\documentclass[journal,twoside,shortpaper,9pt]{IEEEtran}

\usepackage{bm}
\usepackage[usenames,dvipsnames]{xcolor}
\usepackage{srcltx}
\usepackage{psfrag}
\usepackage{epsfig}
\usepackage{graphicx}
\usepackage{epstopdf}
\usepackage{color}
\usepackage[tbtags,sumlimits,nointlimits,reqno]{amsmath}
\usepackage{amssymb}
\usepackage{color}
\usepackage{float}
\usepackage{subfigure}
\usepackage{gensymb}
\usepackage{cuted}
\usepackage{stfloats}
\title{Refracting Metasurfaces without Spurious Diffraction}

\author{Guillaume Lavigne, Karim Achouri, Viktar Asadchy, Sergei Tretyakov, and Christophe Caloz}

\begin{document}

\maketitle

\begin{abstract}
Refraction represents one of the most fundamental operations that may be performed by a metasurface. However, simple phase-gradient metasurface designs suffer from restricted angular deflection due to spurious diffraction orders. It has been recently shown, using a circuit-based approach, that refraction without spurious diffraction, or diffraction-free, can fortunately be achieved by a transverse (or in-plane polarizable) metasurface exhibiting either loss-gain, nonreciprocity or  bianisotropy. Here, we rederive these conditions using a medium-based -- and hence more insightfull -- approach based on Generalized Sheet Transition Conditions (GSTCs) and surface susceptibility tensors, and experimentally demonstrate two diffraction-free refractive metasurfaces that are essentially lossless, passive, bianisotropic and reciprocal.
\end{abstract}

\section{Introduction}\label{sec:intro}

Metasurfaces represent a powerful electromagnetic technology that has experienced spectacular development over the past lustrum~\cite{glybovski2016metasurfaces,urbas2016roadmap,tretyakov2015metasurfaces}. They have already lead to a diversity of applications, including single-layer perfect absorption~\cite{ra2013total}, polarization twisting~\cite{shi2014dual}, power harvesting~\cite{almoneef2015metamaterial}, orbital angular momentum multiplexing~\cite{pfeiffer2014controlling,achouri2016comparison}, spatial processing~\cite{achouri2016metasurface} and flat lensing~\cite{khorasaninejad2016metalenses}, and there seems to be much more to be discovered and developed in this area.

One of the most fundamental operations that a metasurface may perform is generalized refraction and reflection~\cite{yu2011light}, as most metasurface field transformations involve these phenomena. Such operations have been achieved in \emph{blazed gratings}~\cite{neviere1991electromagnetic,destouches200599}, and later in \emph{planar phase-gradient metasurfaces}~\cite{yu2011light}, however with restriction to small angle differences between the incident and refracted or reflected beams and with the presence of spurious diffraction orders~\cite{achouri2016comparison,asadchy2016perfect,estakhri2016wave,epstein2016arbitrary}.

Fortunately, it is possible to achieve refraction without spurious diffraction, that we shall hereafter refer to as \emph{diffraction-free} refraction for short,  by introducing more complexity in the metasurface design. This has been clearly demonstrated in~\cite{asadchy2016perfect}, which shows that such operation may be accomplished by a \emph{transverse}\footnote{A transverse metasurface is a metasurface characterized only by tensor components that are parallel to the plane of the metasurface (i.e. ``in-plane'' or~$\parallel$) or, equivalently, perpendicular to the normal of the metasurface. Including longitudinal (i.e. ``off-plane'' or $\perp$) tensor components, i.e. components in the direction of the normal of the metasurface, immediately brings about much greater complexity because, as shown in~\cite{achouri2014general}, this transforms otherwise algebraic GSTC equations into differential equations. The reduction of the general to a transverse metasurfaces reduces the four bianisotropic constitutive parameters from $3\times 3$ tensors ($4\times 9=36$ elements) to $2\times 2$ tensors ($4\times 4=16$ elements).\label{footnote:transverse_nat}} metasurface if that metasurface exhibits any one of the following three properties:
\begin{enumerate}
  \item monoisotropy with loss and gain~\cite{zhu2015passive},
  \item  nonreciprocity~\cite{asadchy2016perfect}, or
  \item bianisotropy~\cite{wong2016reflectionless,chen2017experimental,lavigne2017perfectly}.
\end{enumerate}

Among these properties, the practically most convenient is certainly the third one, since it allows to achieve diffraction-free generalized refraction with a metasurface that is purely lossless and passive, avoiding complex amplification, and at the same time reciprocal, avoiding non-integrable magnetic materials~\cite{lax1962microwave} or complex magnetless structures~\cite{kodera2011artificial,wang2012gyrotropic,taravati2017nonreciprocal}.

The work reported in~\cite{wong2016reflectionless}, and related experimentation~\cite{chen2017experimental}, represents the first synthesis of a diffraction-free generalized refractive metasurface. In that paper, the authors use a \emph{circuit-based} approach with generalized scattering parameters to match the impedances of the oblique incident and transmitted waves across a \emph{layered} metasurface structure. As a result, they obtain analytical expressions for the admittances of each of the layers constituting the metasurface. Here, as an extension of the short report~\cite{lavigne2017perfectly}, we present a fundamentally different and also more general approach of the same problem. This approach uses surface susceptibilities synthesized~\cite{achouri2015synthesis,achouri2014general,achouri2017mathematical} by Generalized Sheet Transition Conditions (GSTCs)~\cite{idemen2011discontinuities}, and is therefore a \emph{medium-based} rather than a circuit-based approach, which inherently brings about greater insight into the physics of the problem. Moreover, it treats the metasurface as a \emph{global entity}, without any restriction regarding its structure, and may therefore accommodate different implementations, via subsequent scattering parameter mapping~\cite{achouri2014general}. Secondly, starting from a completely general bianisotropic metasurface, this approach naturally reveals the three diffraction-free conditions derived in~\cite{asadchy2016perfect}, and ultimately leads to closed-form expressions for the bianisotropic susceptibility tensors. Finally, we provide an experimentally demonstrate two diffraction-free bianisotropic reciprocal refractive metasurfaces.

The paper is organized as follows. Section~\ref{sec:ref_met_synt} presents the GSTC synthesis of the metasurface susceptibility tensors and discusses the physics of the metasurfaces corresponding to the three above options. Next, Sec.~\ref{sec:scat_par_map} maps the synthesized susceptibilities onto scattering parameters as an intermediate step to discretize the metasurface. Using this mapping, Sec.~\ref{sec:des_scat_part} determines the scattering particles corresponding to each metasurface cell. Simulation and experimental validations are provided in Sec.~\ref{sec:sim_exp}. Finally, conclusions are given in Sec.~\ref{sec:concl}.

\section{Refractive Transverse Metasurface Synthesis}\label{sec:ref_met_synt}
\subsection{Generalized Refraction and GSTC Synthesis}
\begin{figure}[h]
\centering
\includegraphics[width=0.9\columnwidth]{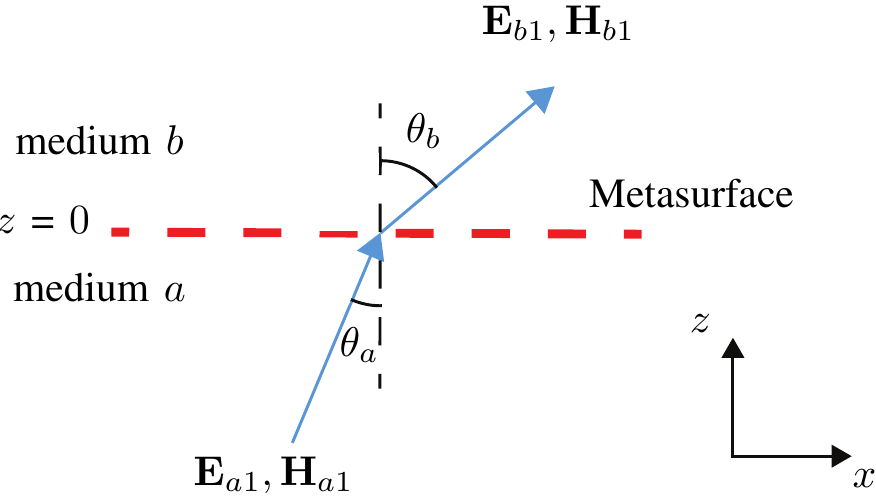}
\caption{Problem of diffraction-free generalized refraction by a metasurface.}\label{fig:direct_t}
\end{figure}

The problem of diffraction-free generalized refraction by a metasurface is represented in Fig.~\ref{fig:direct_t}. The metasurface is placed at $z=0$ in the $xy$-plane of a cartesian coordinate system. We denote $a$ and $b$ the media, possibly having different electromagnetic properties, bounding the metasurface at $z<0$ and $z>0$, respectively. A plane wave, with electric and magnetic fields $\mathbf{E}_{a1}$ and $\mathbf{H}_{a1}$, respectively, impinges from medium $a$ at angle $\theta_a$ onto the metasurface. The metasurface transforms, without any spurious reflection and scattering, this wave into a plane wave, with fields $\mathbf{E}_{b1}$ and $\mathbf{H}_{b1}$, propagating in medium $b$ at angle $\theta_b$.

We shall next perform the synthesis of the diffraction-free generalized refractive metasurface in Fig.~\ref{fig:direct_t} using the GSTC-susceptibility technique presented in~\cite{achouri2014general}. This technique consists in specifying the incident, reflected and transmitted fields and computing the corresponding metasurface susceptibility tensors via GSTCs. According to the definition of diffraction-free generalized refraction, the reflected field will be specified to be zero and the transmitted field will be specified as a pure plane wave without any spurious diffraction orders. In the case of a general transverse bianisotropic metasurface, the GSTCs, defined at $z=0$, are given as
\begin{subequations}\label{eq:gstc}
\begin{equation}
\hat{z} \times \Delta\mathbf{H} = j \omega \epsilon_0 \overline{\overline{ \chi}}_\text{ee} \mathbf{E}_\text{av} +  j \omega \overline{\overline{ \chi}}_\text{em} \sqrt{\mu_0 \epsilon_0}  \mathbf{H}_\text{av} ,
\end{equation}
\begin{equation}
\Delta \mathbf{E} \times \hat{z}   = j \omega \mu_0 \overline{\overline{ \chi}}_\text{mm} \mathbf{H}_\text{av} +  j \omega \mu_0 \overline{\overline{ \chi}}_\text{me} \sqrt{\epsilon_0/\mu_0}  \mathbf{E}_\text{av} ,
\end{equation}
\end{subequations}
where the $\Delta$ symbol and 'av' subscript represent the differences and averages of the tangential electric or magnetic fields on both sides of the metasurface, and $\overline{\overline{ \chi}}_\text{ee}$, $\overline{\overline{ \chi}}_\text{em}$, $\overline{\overline{ \chi}}_\text{me}$, $\overline{\overline{ \chi}}_\text{mm}$ are the bianisotropic susceptibility tensors describing the metasurface. In order to realize the simplest and most fundamental generalized refraction operation, we require the metasurface to be non-gyrotropic so as to avoid polarization alteration. As a result, the s-polarization and p-polarization problems are independent from each other and can therefore be treated separately. We shall therefore, without loss of generally, only treat the p-polarization problem.

\subsection{Monoanisotropic Metasurface}\label{sec:monoanis_ms}

We heuristically start with a metasurface having the simplest constitutive parameters, a \emph{monoanisotropic} metasurface. In the p-polarization case, for which the metasurface appears monoisotropic, the fields corresponding to the scenario of Fig.~\ref{fig:direct_t} are
\begin{subequations}\label{eq:dir_transf}
\begin{equation}
\mathbf{E}_{a1}=(\cos\theta_a\hat{x}+\sin\theta_a\hat{z})e^{-j(k_{ax}x+k_{az}z)},
\mathbf{H}_{a1}=\frac{e^{-jk_{ax}x}}{\eta_a}\hat{y},
\end{equation}
\begin{equation}
\mathbf{E}_{b1}=T_\text{p}(\cos\theta_b\hat{x}+\sin\theta_b\hat{z})e^{-j(k_{bx}x+k_{bz}z)},
\mathbf{H}_{b1}=T_\text{p}\frac{e^{-jk_{bx}x}}{\eta_b}\hat{y},
\end{equation}
\end{subequations}
where $T_\text{p}$ is the (parallel-polarization) transmission coefficient, $\eta_{(a,b)}=\sqrt{\mu_{(a,b)}/\epsilon_{(a,b)}}$, and
\begin{subequations}\label{eq:k_vector}
\begin{equation}
k_{(a,b)x} = k_{a,b} \sin\theta_{a,b},
\end{equation}
\begin{equation}
k_{(a,b)z} = k_{a,b} \cos\theta_{a,b},
\end{equation}
\end{subequations}
where $k_{(a,b)}=\sqrt{\mu_{(a,b)}\epsilon_{(a,b)}}k_0$ with $k_0=\omega/c_0$.

Based on the above assumption of a transverse metasurface, the GSTCs~\eqref{eq:gstc} involve only the $x$ and $y$ components of these fields, evaluated at $z=0$, which, also accounting for the p-polarization monoisotropy, results into
\begin{subequations}\label{eq:2}
\begin{equation}\label{eq:chixxee_mono_transv}
\chi^{{xx}}_\text{ee} = \frac{-\Delta H_{y1}}{j \omega \epsilon_0 E_{x,\text{av}1} },
\end{equation}
\begin{equation}
\chi^{{yy}}_\text{mm} = \frac{-\Delta E_{x1}}{j \omega \mu_0 H_{y,\text{av}1} },
\end{equation}
\end{subequations}
with
\begin{subequations}\label{eq:Dav}
\begin{equation}
\Delta E_{x1}=E_{bx1}-E_{ax1}=T_\text{p}\cos\theta_be^{-jk_{bx}x}-\cos\theta_ae^{-jk_{ax}x},
\end{equation}
\begin{equation}
\Delta H_{y1}=H_{by1}-H_{ay1}=T_\text{p}e^{-jk_{bx}x}/\eta_b-e^{-jk_{ax}x}/\eta_a,
\end{equation}
\begin{equation}
E_{x,\text{av}1}=\frac{E_{ax1}+E_{bx1}}{2}= \frac{\cos\theta_ae^{-jk_{ax}x} + T_\text{p}\cos\theta_be^{-jk_{bx}x}}{2}
\end{equation}
\begin{equation}
H_{y,\text{av}1}=\frac{H_{ay1}+H_{by1}}{2}=\frac{e^{-jk_{ax}x}/\eta_a+T_\text{p}e^{-jk_{bx}x}/\eta_b}{2}
\end{equation}
\end{subequations}
where the subscript `1' has been introduced for later convenience. The only unknown in these relations is the transmission coefficient, $T_\text{p}$. This coefficient may be obtained by enforcing power conservation across the metasurface,
\begin{equation}\label{eq:poynting}
\frac{1}{2}\text{Re}\left((E_{ax\text{1}}\hat{x})\times(H^*_{ay\text{1}}\hat{y})\right)
=\frac{1}{2}\text{Re}\left((E_{bx\text{1}}\hat{x})\times(H^*_{by\text{1}}\hat{y})\right),
\end{equation}
whose resolution for $T_\text{p}$ with~\eqref{eq:dir_transf} yields
\begin{equation}\label{eq:T_coeff}
T_\text{p} = \sqrt{\frac{\eta_b\cos\theta_a}{\eta_a\cos\theta_b}}
\end{equation}
which is thus a fundamental \emph{condition for power conserving diffraction-free refraction}.

Inserting~\eqref{eq:Dav} into~\eqref{eq:2} yields the periodic complex susceptibility functions
\begin{subequations}\label{eq:Imag_X}
\begin{equation}\label{eq:Re_Xeexx_mono_trans}
\begin{split}
\text{Re}(\chi^{{xx}}_\text{ee}) = \frac{-2 k_a k_b T_\text{p} (\eta_a k_b k_{az}+\eta_b k_a k_{bz}) \sin (\alpha x)}{\epsilon_0 \omega \eta_a \eta_b (k_b^2 k_{az}^2 + k_a^2 k_{bz}^2 T_\text{p}^2 + 2k_a k_{az} k_b k_{bz} T_\text{p} \cos(\alpha x)  )},
\end{split}
\end{equation}
\begin{equation}\label{eq:Im_Xeexx_mono_trans}
\begin{split}
&\text{Im}(\chi^{{xx}}_\text{ee}) = \\
& \frac{2k_a k_b(\eta_a k_a k_{bz} T_\text{p}^2 - \eta_b k_{az} k_b +T_\text{p}(\eta_a k_b k_{az}-\eta_b k_a k_{bz})\cos(\alpha x) ))}{\epsilon_0 \omega \eta_a \eta_b (k_b^2 k_{az}^2 + k_a^2 k_{bz}^2 T_\text{p}^2 + 2k_a k_{az} k_b k_{bz} T_\text{p} \cos(\alpha x))},
\end{split}
\end{equation}
\begin{equation}
\text{Re}(\chi^{{yy}}_\text{mm}) = \frac{-2 \eta_a \eta_b(\eta_a k_{az} k_b + \eta_b k_a k_{bz})T_\text{p}\sin (\alpha x)}{k_a k_b \mu_0 \omega(\eta_b^2 \eta_a^2 T_\text{p}^2 + 2\eta_a \eta_b T_\text{p})\cos (\alpha x)},
\end{equation}
\begin{equation}
\begin{split}
&\text{Im}(\chi^{{yy}}_\text{mm}) = \\
& \frac{2 \eta_a \eta_b( \eta_a k_a k_{bz} T_\text{p}^2-\eta_b k_{az} k_b+ (\eta_a k_a k_{bz} T_\text{p}^2 - \eta_b k_b k_{az})T_\text{p}^2 \cos(\alpha x))}{k_a k_b \mu_0 \omega(\eta_b^2 \eta_a^2 T_\text{p}^2 + 2\eta_a \eta_b T_\text{p})\cos (\alpha x)},
\end{split}
\end{equation}
\end{subequations}
with $\alpha = k_{ax}-k_{bx}$. Plots of these functions may be found in~\cite{achouri2016comparison}. The non-zero imaginary parts of $\chi^{{xx}}_\text{ee}$ and $\chi^{{yy}}_\text{mm}$, tensorially corresponding to the loss and gain relations
$\overline{\overline{ \chi}}_\text{ee}^T\neq\overline{\overline{\chi}}_\text{ee}^*$ and $\overline{\overline{ \chi}}_\text{mm}^T\neq\overline{\overline{\chi}}_\text{mm}^*$, where the superscripts $T$ and $*$ denote the transpose and conjugate operation respectively~\cite{rothwell2008electromagnetics}, indicate the presence of loss (negative imaginary part) and gain (positive imaginary part) alternating along the metasurface. This synthesis corresponds to the first way of obtaining a diffraction-free refractive metasurface, as shown in~\cite{asadchy2016perfect}.

\subsection{Bianisotropic Metasurface}\label{sec:bianis_meta}

Since specifying a monoanisotropic (or monoisotropic) metasurface leads only to the loss and gain option for diffraction-free refraction, as just found, complexity must be added to the metasurface to obtain the nonreciprocity and bianisotropy options. The non-gyrotropy assumption requires $\chi^{xy}_\text{ee,mm}=\chi^{yx}_\text{ee,mm}=\chi^{xx}_\text{em,me}=\chi^{yy}_\text{em,me}=0$, and hence eliminates 8 of the 16 terms of a transverse metasurface, and, among the remaining 8 terms, 4 are for p-polarization and 4 are for s-polarization. Therefore, still assuming p-polarization, only the \emph{bianisotropic} two terms $\chi^{xy}_\text{em}$ and $\chi^{yx}_\text{me}$ can be added to $\chi^{xx}_\text{ee}$ and $\chi^{yy}_\text{mm}$. This 4-element susceptibility set allows for two fundamentally new possibilities: a)~$\chi^{xy}_\text{em}\neq-\chi^{yx}_\text{me}$, and b)~$\chi^{xy}_\text{em}=-\chi^{yx}_\text{me}$. The latter tensorially generalizes to $\overline{\overline{ \chi}}_\text{em}=-\overline{\overline{ \chi}}_\text{me}^T$, where the superscript `$T$' represents the transpose operation, which is the only condition for \emph{reciprocity} in the prevailing non-gyrotropic situation~\cite{rothwell2008electromagnetics}, and the former corresponds thus to a \emph{nonreciprocal} metasurface. These two possibilities correspond to options 2) and 3), respectively, in~\cite{asadchy2016perfect}. In each of the two cases, one has to describe the phenomenon (reciprocity or nonreciprocity) by also specifying the transformation in the reverse direction, namely the direction from medium $b$ to medium $a$, which brings about two additional equations, leading to a full-rank matrix system of order 4.

\begin{figure}[h]
\centering
\includegraphics[width=0.9
\columnwidth]{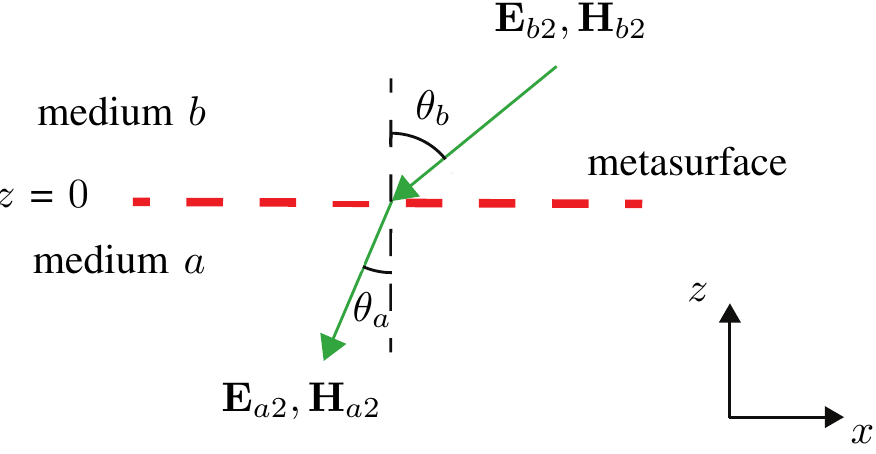}
\caption{Reverse transformation reciprocal to that of Fig.~\ref{fig:direct_t}.}\label{fig:recip_t}
\end{figure}

\begin{figure}[h]
\centering
\includegraphics[width=1.1\columnwidth]{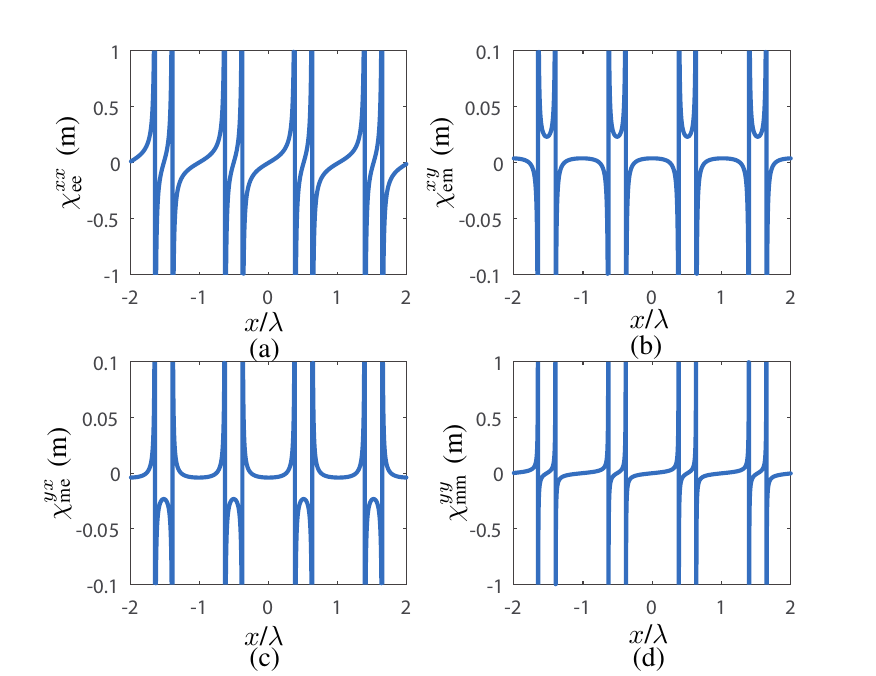}
\caption{Susceptibility functions~\eqref{eq:biani_X} for $\theta_a=0^\circ$ and $\theta_b=70^\circ$. (a)~$\chi^{xx}_\text{ee}$ (purely real). (b)~$\chi^{xy}_\text{em}$ (purely imaginary). (c)~$\chi^{yx}_\text{me}$  (purely imaginary). (d)~$\chi^{yy}_\text{mm}$  (purely real). }\label{fig:plot_X_biani}
\end{figure}

In the nonreciprocal case, one may specify any reverse transformation, such as for instance refraction in different directions or absorption. However, as mentioned in Sec.~\ref{sec:intro}, we are primarily interested here in realizing a reciprocal metasurface. The corresponding reverse transformation, involving the same angles as in Fig.~\ref{fig:direct_t}, is shown in Fig.~\ref{fig:recip_t}.
Thus, the proper synthesis equations, including both the reciprocal direct (subscript `1') and reverse (subscript `2') transformations, may be compactly written as
\begin{equation}\label{eq:two_transf_syst}
\begin{bmatrix}
    \Delta H_{y1}    &      \Delta H_{y2} \\
    \Delta E_{x1}     &      \Delta E_{x2} \\
\end{bmatrix}
=
\begin{bmatrix}
    -j\omega\epsilon_0\chi^{xx}_\text{ee}      & -jk_0\chi^{xy}_\text{em}  \\
    -jk_0\chi^{yx}_\text{me}       & -j\omega\mu_0\chi^{yy}_\text{mm} \\
\end{bmatrix}
\begin{bmatrix}
          E_{x1,\text{av}}      &   E_{x2,\text{av}} \\
      H_{y1,\text{av}} & H_{y2,\text{av}}
\end{bmatrix},
\end{equation}
whose first and second columns correspond to the direct and reverse transformations, respectively. The fields corresponding to the reverse transformation [Fig.~\ref{fig:recip_t}] read
\begin{subequations}\label{eq:rec_transf}
\begin{equation}
\mathbf{E}_{a2}=-(\cos\theta_a\hat{x}+\sin\theta_a\hat{z})e^{j(k_{ax}x+k_{az}z)},
\mathbf{H}_{a2}=\frac{e^{jk_{ax}x}}{\eta_a}\hat{y},
\end{equation}
\begin{equation}
\mathbf{E}_{b2}=-T_\text{p}(\cos\theta_b\hat{x}+\sin\theta_b\hat{z})e^{j(k_{bx}x+k_{bz}z)},
\mathbf{H}_{b2}=T_\text{p}\frac{e^{jk_{bx}x}}{\eta_b}\hat{y},
\end{equation}
\end{subequations}
corresponding to
\begin{subequations}\label{eq:Dav2}
\begin{equation}
\Delta E_{x2}=E_{bx2}-E_{ax2}= -T_\text{p}\cos\theta_be^{jk_{bx}x}+\cos\theta_ae^{jk_{ax}x},
\end{equation}
\begin{equation}
\Delta H_{y2}=H_{by2}-H_{ay2}= T_\text{p}e^{jk_{bx}x}/\eta_b-e^{jk_{ax}x}/\eta_a,
\end{equation}
\begin{equation}
E_{x,\text{av}2}=\frac{E_{ax2}+E_{bx2}}{2}=\frac{-\cos\theta_ae^{jk_{ax}x}- T_\text{p}\cos\theta_be^{jk_{bx}x}}{2},
\end{equation}
\begin{equation}
H_{y,\text{av}2}=\frac{H_{ay2}+H_{by2}}{2}=\frac{e^{jk_{ax}x}/\eta_a+T_\text{p}e^{jk_{bx}x}/\eta_b}{2}.
\end{equation}
\end{subequations}
Inserting~\eqref{eq:Dav} and~\eqref{eq:Dav2} into~\eqref{eq:two_transf_syst} finally yields the sought after transverse susceptiblity functions
\begin{subequations}\label{eq:biani_X}
\begin{equation}
\begin{split}
&\chi^{xx}_\text{ee} = \\
&\frac{-4 k_a k_b  T_\text{p} \sin (\alpha x))}{\epsilon_0 \omega \left(T_\text{p} (\eta_a k_b k_{az}+\eta_b k_a k_{bz}) \cos (\alpha x))+\eta_a k_b k_{az}+\eta_b k_a k_{bz} T_\text{p}^2\right)},
\end{split}
\end{equation}

\begin{equation}
\begin{split}
&\chi^{xy}_\text{em} =\\
&\frac{2 j \left(T_\text{p} (\eta_a k_b k_{az}- \eta_b k_a k_{bz}) \cos (\alpha x))-\eta_b k_b k_{az}+\eta_a k_a k_{bz} T_\text{p}^2\right)}{k_0 \left(T_\text{p} (\eta_a k_b k_{az}+\eta_b k_a k_{bz}) \cos (\alpha x))+\eta_a k_b k_{az}+\eta_b k_a k_{bz} T_\text{p}^2\right)},
\end{split}
\end{equation}

\begin{equation}
\begin{split}
&\chi^{yx}_\text{me} =\\
&\frac{2 j \left(T_\text{p} (\eta_b k_a k_{bz}-\eta_a k_b k_{az} ) \cos (\alpha x))-\eta_b k_b k_{az}+\eta_a k_a k_{bz} T_\text{p}^2\right)}{k_0 \left(T_\text{p} (\eta_a k_b k_{az}+\eta_b k_a k_{bz}) \cos (\alpha x))+\eta_a k_b k_{az}+\eta_b k_a k_{bz} T_\text{p}^2\right)},
\end{split}
\end{equation}

\begin{equation}
\begin{split}
&\chi^{yy}_\text{mm} = \\
&\frac{-4 \eta_a \eta_b k_{az} k_{bz}  T_\text{p} \sin (\alpha x))}{\mu_0 \omega \left(T_\text{p} (\eta_a k_b k_{az}+\eta_b k_a k_{bz}) \cos (\alpha x))+\eta_a k_b k_{az}+\eta_b k_a k_{bz} T_\text{p}^2\right)},
\end{split}
\end{equation}

\end{subequations}

with $\alpha = k_{ax}-k_{bx}$. These relations are plotted in Fig.~\ref{fig:plot_X_biani} for $\theta_a=0^\circ$ and $\theta_b=70^\circ$, and considering air on both sides of the metasurface.

\subsection{Properties of the Synthesized Metasurface}\label{sec:properties_ms}

The metasurface characterized by the susceptibilities in~\eqref{eq:biani_X} possesses the following properties:

\begin{itemize}
 \item  It is \emph{bianisotropic}, as already noted in Sec.~\ref{sec:bianis_meta}, since $\chi^{xy}_\text{em}\neq 0$ and $\chi^{yx}_\text{me}\neq 0$.
  \item  As a result of bianisotropy, it is \emph{asymmetric}, as will be shown in the corresponding scattering parameters to be given in Sec.~\ref{sec:scat_par_map}.
  \item It is \emph{reciprocal}, as noted in Sec.~\ref{sec:bianis_meta}, since $\chi^{xy}_{\text{em}} = - \chi^{yx}_{\text{me}}$, which is equivalent to $T_\text{p}=\sqrt{\eta_b \cos \theta_a / \eta_a \cos \theta_b}$ (power conservation), tensorially corresponding to the relation $\overline{\overline{ \chi}}_\text{me}^T =-\overline{\overline{\chi}}_\text{em}$.
  \item It is \emph{passive} and \emph{lossless}, since $\chi^{xx}_\text{ee},\chi^{yy}_\text{mm}\in\mathbb{R}$ and $\chi^{xy}_\text{em},\chi^{yx}_\text{me}\in\mathbb{I}$, tensorially corresponding to the relations
$\overline{\overline{ \chi}}_\text{ee}^T =\overline{\overline{\chi}}_\text{ee}^*$, $\overline{\overline{ \chi}}_\text{mm}^T = \overline{\overline{\chi}}_\text{mm}^*$ and
$\overline{\overline{ \chi}}_\text{me}^T\ =\overline{\overline{\chi}}_\text{em}^*$~\cite{rothwell2008electromagnetics}.
  \item It is \emph{periodic} in $x$ with period $k_{ax}-k_{bx}$, as seen in Fig.~\ref{fig:plot_X_biani}, corresponding to the periodic field momentum transformation~\cite{salem2014manipulating} operated by the metasurface.
  \item Since it is periodic, it inherently supports an \emph{infinite number of space harmonics}~\cite{ishimaru1991electromagnetic}. From the fact that the metasurface is synthesized so as to scatter only in one direction, all the space harmonics (or diffraction orders) \emph{other than that corresponding to the specified refraction angle} must be \emph{evanescent} (i.e. transformed to leaky or surface waves).
\end{itemize}

\section{Scattering Parameter Mapping}\label{sec:scat_par_map}

The task now is to establish a proper link between the mathematical transverse susceptibility functions~\eqref{eq:biani_X} and the corresponding real metasurface, composed of an array of scattering particles. Specifically, this includes discretizing these susceptibility functions in subwavelength cells and determining the appropriate particle geometries for all the cells.

To build a metasurface with the assumed purely transverse susceptibility, $\overline{\overline{\chi}}_\text{t}$, we shall proceed as follows:
\begin{enumerate}
  \item Map the synthesized susceptibility parameters onto \emph{normal-incidence} scattering parameters. The reason to use normal incidence \emph{in the design procedure} is twofold. First, this is a necessary condition to ensure that only the \emph{transverse} terms of the particle susceptibilities or, more precisely, polarizabilities, get excited\footnote{In the most general case of full-tensor bianisotropic metasurface, one would have to realize~36 parameters (Footnote~\ref{footnote:transverse_nat}). Given that GSTCs are composed of 4 (transverse) equations, such a design would in principle demand to specify $36/4=9$ transformations across the unit cell, including oblique incidence, which would be prohibitively complicated. Restricting the incidence to be normal, we ensure to select out and realize only the 16~transverse susceptibilities, which corresponds to $16/4=4$ transformations, 2 for each polarization (p and s) and, for each polarization, 2 for each incidence side (Fig.~\ref{fig:direct_t} and Fig.~\ref{fig:recip_t}). This eventually requires only a simple normal-incidence scattering simulation for each $z$-direction.}. Second, this will lead to the \emph{simplest} possible simulation set-up for each specific particle (step 3 below), obliqueness being produced by the phase gradient between cells. For p-polarization, and assuming that the metasurface is reciprocal and surrounded by air, the corresponding relations are found, following the procedure in~\cite{achouri2014general}, as
\begin{subequations}\label{eq:S_param_from_X}
\begin{equation}\label{eq:S11_param_from_X}
S^{xx}_{11} =\frac{-2j\left( 2k_0\eta_0 \chi_{\text{me}}^{yx} +\mu_0 \omega \chi_{\text{mm}}^{yy} - \eta_0^2 \epsilon_0 \omega \chi_{\text{ee}}^{xx} \right)}{D^{xx}},
\end{equation}
\begin{equation}\label{eq:S22_param_from_X}
S^{xx}_{22} = \frac{2j\left( 2k_0\eta_0 \chi_{\text{me}}^{yx} -\mu_0 \omega \chi_{\text{mm}}^{yy} + \eta_0^2 \epsilon_0 \omega \chi_{\text{ee}}^{xx} \right)}{D^{xx}},
\end{equation}
\begin{equation}
S^{xx}_{21} = S^{xx}_{12} = \frac{-j \eta_0 \left(4+ k_0^2 (\chi_{\text{me}}^{yx})^2+ \mu_0 \epsilon_0 \omega^2 \chi_{\text{ee}}^{xx} \chi_{\text{mm}}^{yy}  \right)}{D^{xx}}.
\end{equation}
with
\begin{equation}
\begin{split}
D^{xx}
=&-2j\mu_0 \omega\chi_{\text{mn}}^{yy}\\
&+\eta_0\left(-4 + k_0^2 (\chi_{\text{me}}^{yx})^2 + \epsilon \omega \chi_{\text{ee}}^{xx}\left(-2j \eta_0 + \mu_0 \omega \chi_{\text{mm}}^{yy} \right) \right),
\end{split}
\end{equation}
\end{subequations}
where the $x$'s in the superscript $xx$ correspond to the transverse component of the p-polarized fields [Eqs.~\eqref{eq:dir_transf} and~\eqref{eq:rec_transf}], assuming also non-gyrotropy. We have thus obtained the scattering matrix periodic functions $\overline{\overline{S}}^{_{xx}}(x,y)$ corresponding to $\overline{\overline{\chi}}_\parallel(x,y)$. As announced in Sec.~\ref{sec:properties_ms}, Eqs.~\eqref{eq:S11_param_from_X} and~\eqref{eq:S11_param_from_X} reveal that the metasurface is asymmetric, since $S^{xx}_{11}\neq S^{xx}_{22}$.
\item Discretize the periodic functions~\eqref{eq:S_param_from_X}, or $\overline{\overline{S}}^{_{xx}}(x,y)$, in subwavelength cells in order to ensure their safe sampling in terms of Nyquist theorem. This leads to  the discrete function $\overline{\overline{S}}^{_{xx}}(x_i,y_j)$, for $i,\ldots,N_x$ and $j,\ldots,N_y$, where $N_x$ and $N_y$ represent the number of cells along the $x$ and $y$ directions, respectively.
  \item Select a generic particle structure and geometry that may be adjusted to cover the phase and amplitude range of $\overline{\overline{S}}^{_{xx}}(x_i,y_j)$ across the entire metasurface. For simplicity and computational efficiency, compute the scattering parameters (under normal incidence) of each cell separately and within \emph{periodic boundary conditions}. Even though the final metasurface will be locally aperiodic, i.e. made of different adjacent cells, periodic boundary conditions will reasonably approximate the coupling to slightly different neighbours.
  \item Since $\overline{\overline{S}}^{_{xx}}(x_i,y_j)$ is periodic, the period includes the complete set of all the different cells, and the overall structure will consist in the periodic repetition of the corresponding \emph{super-cell}. Now, simulate this supercell within periodic boundary conditions with the specified incidence angle, and optimize the geometry of the particles so as to maximize energy refraction in the specified direction, i.e., specifically, in the proper diffraction order corresponding to the supercell.
\end{enumerate}

Note that it is practically difficult to realize scattering particles with purely transverse polarizability and hence purely transverse susceptibility. Practical metasurfaces typically always include small non-zero longitudinal susceptibility terms, $\overline{\overline{\chi}}_\perp(x,y)$. Such terms are not excited in 3) above, due to normal incidence, but would play a role in 4) above, given the oblique angle\footnote{The final design may thus, rigorously, include a small $\overline{\overline{\chi}}_\perp(x,y)$ associated with a slightly modified $\overline{\overline{\chi}}_\parallel(x,y)$. Such a metasurface would strictly correspond to a diffraction-free-refraction design different than the initially purely transverse one. This does not represent any contradiction since the synthesis corresponds to an \emph{inverse problem}, naturally admitting multiple solutions.}. We shall select a generic particle without longitudinal metallizations, to avoid strong perpendicular electric moments, and without transverse loops, to avoid strong perpendicular magnetic moments. We may therefore expect negligible $\overline{\overline{\chi}}_\perp(x,y)$ and a design essentially correspond to the assumed purely transverse one.

\section{Design of Scattering Particles}\label{sec:des_scat_part}

We shall design here two diffraction-free refractive transverse metasurfaces to illustrate the theory of the previous sections: the first metasurface with $(\theta_a,\theta_b)=(20^\circ,-28^\circ)$ at 10~GHz and the second with $(\theta_a,\theta_b)=(0^\circ,-70^\circ)$ at 10.5~GHz. For this purpose, we shall follow the procedure described in Sec.~\ref{sec:scat_par_map}. For experimental simplicity, we assume that the metasurface is entirely surrounded by free space ($k_a = k_b = k_0$, $\eta_a = \eta_b = \eta_0$).

As the step 1), we insert the susceptibilities given by~\eqref{eq:biani_X}, with $k_a = k_b = k_0$, into~\eqref{eq:S_param_from_X}. This yields the scattering parameter functions
\begin{subequations}\label{eq:s-param_normal}
\begin{equation}
\begin{split}
&S^{xx}_\text{11} = \\
&\frac{(-k_0^2+k_{az}k_{bz})\sin[(k_{ax}-k_{bx})x]+j k_0 (k_{az}-k_{bz}) \cos[(k_{ax}-k_{bx})x]}{(k_0^2+k_{az}k_{bz})\sin[(k_{ax}-k_bx)x]+j k_0 (k_{az}+k_{bz}) \cos[(k_{ax}-k_{bx})x]},
\end{split}
\end{equation}
\begin{equation}
\begin{split}
&S^{xx}_\text{22} = \\
&\frac{(-k_0^2+k_{az}k_{bz})\sin[(k_{ax}-k_{bx})x]-j k_0 (k_{az}-k_{bz}) \cos[(k_{ax}-k_{bx})x]}{(k_0^2+k_{az}k_{bz})\sin[(k_{ax}-k_bx)x]+j k_0 (k_{az}+k_{bz}) \cos[(k_{ax}-k_{bx})x]},
\end{split}
\end{equation}
\begin{equation}
\begin{split}
&S^{xx}_\text{12} = S^{xx}_\text{21} = \\
& \frac{2 j k_0 \sqrt{k_{az} k_{bz}}}{(k_0^2+k_{az}k_{bz})\sin[(k_{ax}-k_bx)x]+j k_0 (k_{az}+k_{bz}) \cos[(k_{ax}-k_{bx})x]}.
\end{split}
\end{equation}
\end{subequations}
It may a priori seem contradictory with the initial assumption of reflection-less refraction to obtain $S^{xx}_\text{11}\neq 0$ and $S^{xx}_\text{22}\neq 0$. However, there is no contradiction if one recalls that Eqs.~\eqref{eq:S_param_from_X} are associated in the \emph{design procedure} with \emph{normal incidence}, both to isolate out transverse susceptibility components and to simulate the cells one by one, whereas the specified incidence angle is generally nonzero. When excited under the specified oblique incidence angle, the metasurface realized by this design methodology will naturally be reflection-less. Note that the metasurface asymmetry predicted in Sec~\ref{sec:properties_ms} is still clearly apparent from the fact that $S^{xx}_{11} \neq S^{xx}_{22}$, since asymmetry for normal incidence implies asymmetry.

% No longer apparent with non-normal expressions
%From~\eqref{eq:s-param_normal}, we can observe that the scattering parameter functions differ more and more from a simple gradient metasurface (which would have $S^{xx}_\text{11}=S^{xx}_\text{2}=0$ and $S^{xx}_\text{12} = S^{xx}_\text{21} = 1 e^{-j k x \sin(\theta_b)}$) as the angle of refraction increases. As a result, we will want to implement refractive metasurfaces for high angle of refraction.

As step 2), we discretize each of the two metasurfaces in 6 different unit cells of size $6 \times 6$~mm  ($\sim \lambda_0/5$) for the metasurface with $(\theta_a,\theta_b)=(20^\circ,-28^\circ)$ and  $5.1 \times 5.1$~mm  ($\sim \lambda_0$/5.6) for the metasurface with $(\theta_a,\theta_b)=(0^\circ,-70^\circ)$.

\begin{figure}[h]
\centering
\includegraphics[width=0.9\columnwidth]{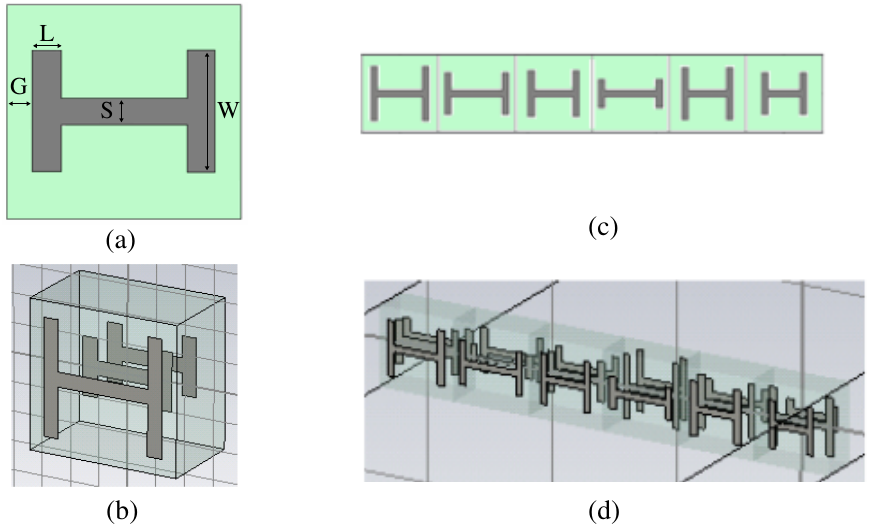}
\caption{Metasurface scattering particles. (a)~Unit cell front view with dog-bone shaped metallic particle. (b)~Unit cell perspective view with dielectric substrates made transparent for visualisation. (c)~Supercell composed of 6 unit cells, front view. (d)~Supercell perspective view.}\label{fig:cells}
\end{figure}

As step 3), we choose scattering particles made of three dog-bone shaped metallic layers separated by 1.52~mm-thick ($\thicksim\lambda_0/20=\lambda_\text{d}/11.55$) Rogers 3003 ($\epsilon_\text{r}=3$, $\tan\delta = 0.0013$) dielectric slabs. The generic dog-bone metallization is shown in Fig.~\ref{fig:cells}~(a), while Fig.~\ref{fig:cells}~(b) shows the corresponding three-layer unit cell. Each unit cell is then optimized with periodic conditions using a commercial software (CST Studio 2014), which provides a reasonable initial guess for the geometry of the dog-bone patterns.

As step 4), we combine the six different unit cells into a supercell, which is periodically repeated to form the whole metasurface. Figure~\ref{fig:cells}~(c) and (d) show the generic structure of the supercell. Finally, the supercell, automatically taking into account the exact (as opposed to periodic) coupling between adjacent unit cells, is optimized. Specifically, we simulate the Floquet space harmonics of the supercell and adjust the geometrical parameters so as to maximize the energy in the desired mode. The dimensions (in mm) of the metallic structures of the different unit cells are listed in Tabs.~\ref{tab:dim_60} and~\ref{tab:dim_80} for the two metasurfaces.

\begin{table}
\caption{Dimensions of the unit cells for the $(\theta_a,\theta_b)=(20^\circ,-28^\circ)$ metasurface.}  \label{tab:dim_60}
\begin{center}
\begin{tabular}{ c c c c c c }
\hline
    &   & W & L & G & S \\ \hline
    & Layer 1 & 4.25 & 0.5 & 0.75 & 0.5 \\
 Cell 1 & Layer 2 & 3.125 &0.5	&0.5	&0.875 \\
  & Layer 3 & 2.5&	0.5&	0.25&	0.5\\ \hline
    & Layer 1 & 3.25&	0.5&	0.5&	0.5 \\
 Cell 2 & Layer 2 & 1.75&	0.5&	0.5&	0.25 \\
  & Layer 3 & 3.25&	0.5&	0.5 & 0.5 \\ \hline
    & Layer 1 & 3.625&	0.5&	1&	0.5 \\
 Cell 3 & Layer 2 & 3&	0.5&	0.75&	0.5 \\
  & Layer 3 & 2.625&	0.5&	0.5&	0.5 \\ \hline
    & Layer 1 & 2.25&	0.5&	0.5&	0.5 \\
 Cell 4 & Layer 2 & 4.75&	0.5&	0.75&	0.5 \\
  & Layer 3 & 4.5&	0.5&	1&	0.5 \\ \hline
    & Layer 1 & 4.25&	0.5&	1&	0.5 \\
 Cell 5 & Layer 2 & 4.25&	0.5&	3.75&	0.5\\
  & Layer 3 &2.375&	0.5&	0.5&	0.5\\ \hline
    & Layer 1 &3&	0.5&	1.25&	0.5 \\
 Cell 6 & Layer 2 &4.125&	0.5&	0.5&	0.5 \\
  & Layer 3 & 1.5&	0.5&	1&	0.5 \\ \hline

\end{tabular}
\end{center}
\end{table}

\begin{table}
\caption{Dimensions of the unit cells for the $(\theta_a,\theta_b)=(0^\circ,-70^\circ)$ metasurface.}\label{tab:dim_80}
\begin{center}
\begin{tabular}{ c c c c c c }
\hline
  &   & W & L & G & S \\ \hline
    & Layer 1 & 3 &	0.5 &	0.375 &	0.5 \\
 Cell 1 & Layer 2 & 2.5	 &0.5	 &0.375	 &0.5 \\
  & Layer 3 & 4	 &0.5	 &0.25	 &0.5 \\ \hline
    & Layer 1 & 3.25	 &0.5	 &0.25	 &0.5\\
 Cell 2 & Layer 2 & 3	 &0.5	 &0.5	 &0.5 \\
  & Layer 3 & 2.5	 &0.5	 &0.25	 &0.5 \\ \hline
    & Layer 1 & 4	 &0.5	 &0.75	 &0.5\\
 Cell 3 & Layer 2 & 3.75	 &0.5	 &0.75	 &0.5 \\
  & Layer 3 & 2.5	 &0.5	 &0.5	 &0.5 \\ \hline
    & Layer 1 & 3.25	 &0.5	 &0.625	 &0.5 \\
 Cell 4 & Layer 2 & 1.5	 &0.5	 &1	 &0.5 \\
  & Layer 3 & 2	 &0.5	 &0.875	 &0.5 \\ \hline
    & Layer 1 & 4.5	 &0.5	 &0.75	 &0.5 \\
 Cell 5 & Layer 2 &4.5	 &0.5	 &0.625	 &0.5\\
  & Layer 3 & 4.25	 &0.5	 &1	 &0.5\\ \hline
    & Layer 1 & 3.25	 &0.5	 &0.875	 &0.5\\
 Cell 6 & Layer 2 & 4.25	 &0.5	 &0.5	 &0.5 \\
  & Layer 3 & 4	 &0.5	 &1	 &0.5 \\ \hline

\end{tabular}
\end{center}
\end{table}

\section{Simulation and Experiment}\label{sec:sim_exp}

\begin{figure}[h]
\centering
\includegraphics[width=0.9
\columnwidth]{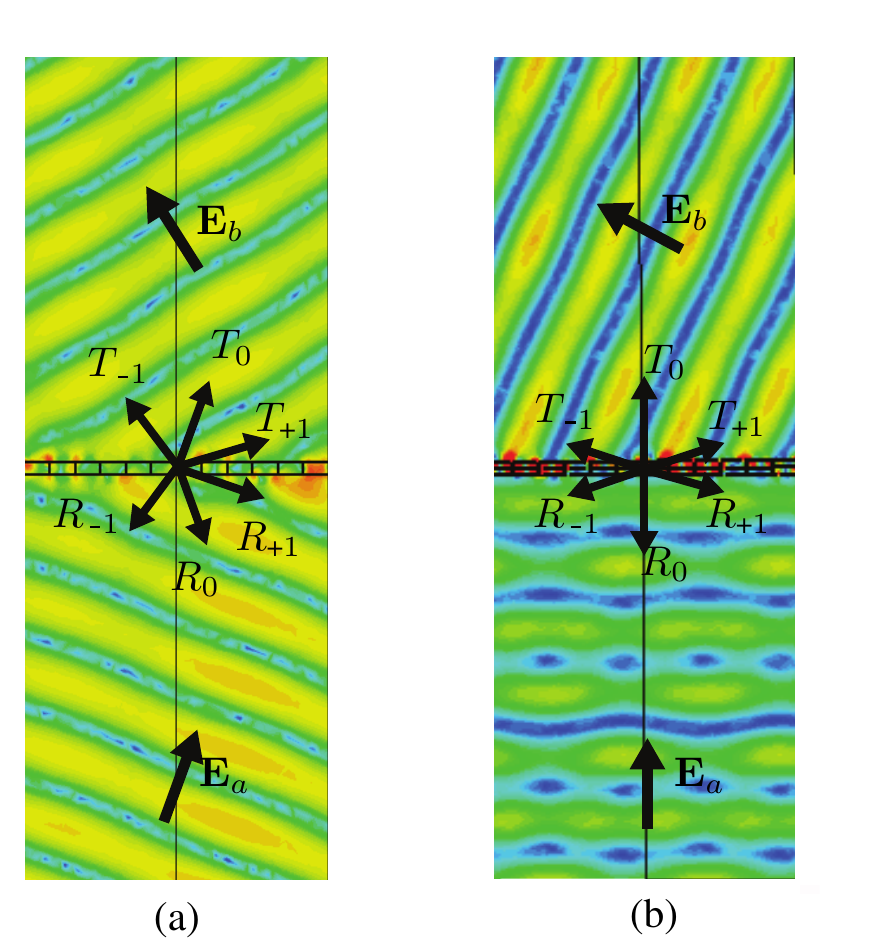}
\caption{Full-wave simulated electric field magnitude for the two diffraction-free bianisotropic reciprocal refractive metasurfaces. The horizontal extent of each figure corresponds to two supercells. The propagating field includes the reflected ($R$) and transmitted ($T$) space harmonics $m=0$, $m=-1$ and $m=+1$, whose directions are indicated by the arrows in the center, while the other space harmonics are evanescent. (a)~Metasurface with $(\theta_a,\theta_b)=(20^\circ,-28^\circ)$ at 10~GHz. (b)~Metasurface with $(\theta_a,\theta_b)=(0^\circ,-70^\circ)$ at 10.5~GHz.}\label{fig:full_wave}
\end{figure}

The full-wave simulated fields of the two diffraction-free bianisotropic reciprocal refractive metasurfaces are plotted in Fig.~\ref{fig:full_wave}. Being perfectly periodic, the metasurface supports in principle an infinite number of space harmonics, as mentioned in the last item of Sec.~\ref{sec:properties_ms}. In both designs, only the space harmonics $m=0$, $m=-1$ and $m=+1$ are propagating, while the others are evanescent, and the incident and refracted waves correspond to the space harmonics $m=0$ and $m=-1$, respectively. Ideally, from synthesis, 100$\%$ of the scattered power should reside in the $m=-1$ space harmonic. Practically, the harmonics $R_0$, $R_{-1}$, $R_{+1}$, $T_0$ and $T_{+1}$ are also weakly excited, due to the imperfections of the metasurface associated with discretization and fabrication restrictions (essentially limited resolution of the metallic particles), already taken into account at this simulation stage.

\begin{figure}[h]
\centering
\includegraphics[width=1.1
\columnwidth]{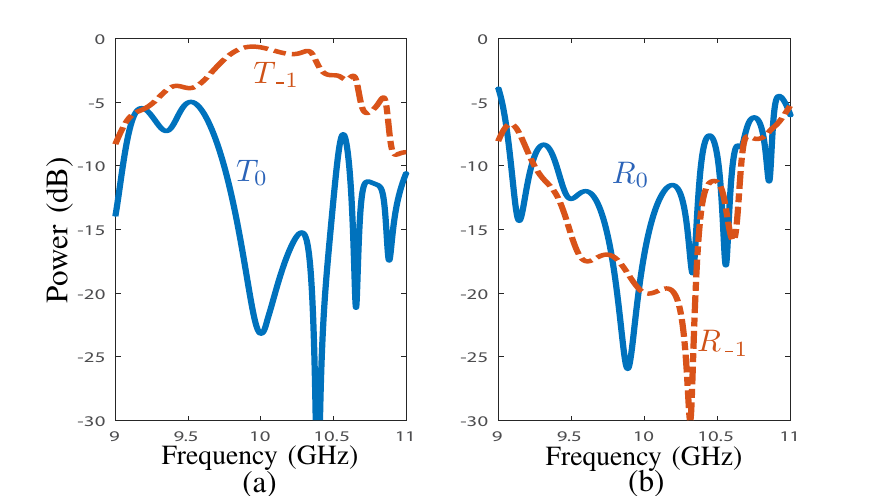}
\caption{Scattering parameters of the $(\theta_a,\theta_b)=(20^\circ,-28^\circ)$ 10~GHz metasurface. (a)~Transmitted propagating space harmonics. (b)~Reflected propagating space harmonics. The $T_{+1}$ and $R_{+1}$ harmonics are also propagating but not visible in these graphs as their magnitudes are lower than -80~dB.}\label{fig:s-param_60}
\end{figure}

\begin{figure}[h]
\centering
\includegraphics[width=1.1
\columnwidth]{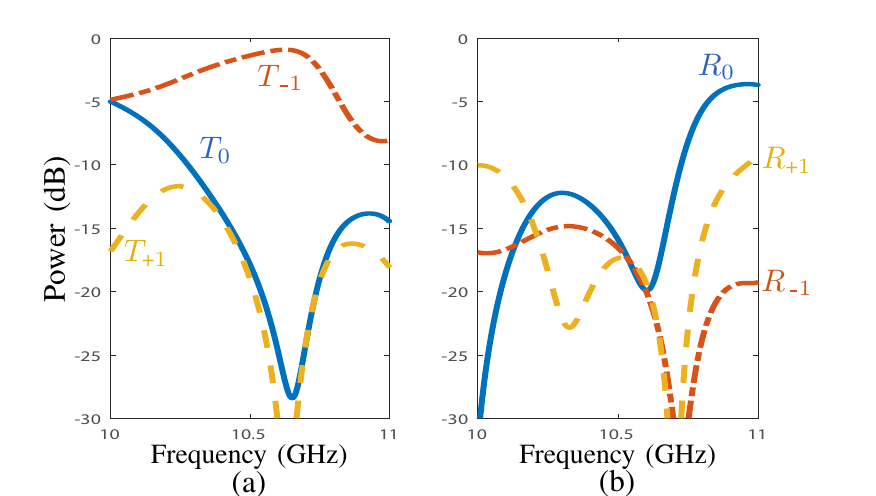}
\caption{Scattering parameters of the $(\theta_a,\theta_b)=(0^\circ,-70^\circ)$ 10.5~GHz metasurface. (a)~Transmitted propagating space harmonics. (b)~Reflected propagating space harmonics. (Note that the optimum has been up-shifted by around 0.1~GHz in the optimization procedure.)}\label{fig:s-param_80}
\end{figure}

The corresponding scattering parameter simulations are shown in Fig.~\ref{fig:s-param_60} and Fig.~\ref{fig:s-param_80} for the $(\theta_a,\theta_b)=(20^\circ,-28^\circ)$ metasurface and the $(\theta_a,\theta_b)=(0^\circ,-70^\circ)$ metasurface, respectively. As expected from synthesis, most of the incident power, except for small conducting and dielectric dissipation loss and negligible coupling to undesired space harmonics, is refracted to the specified direction ($\sim -0.6$~dB for the $(\theta_a,\theta_b)=(20^\circ,-28^\circ)$ metasurface and $\sim -0.9$~dB for the $(\theta_a,\theta_b)=(0^\circ,-70^\circ)$ metasurface) with low reflection ($<-15$~dB).

\begin{figure}[h]
\centering
\includegraphics[width=0.9
\columnwidth]{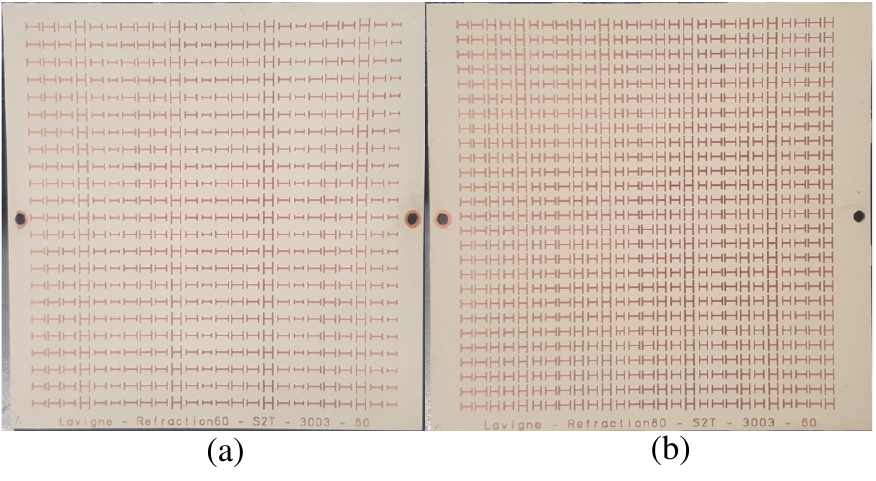}
\caption{Photographs of the two fabricated metasurfaces. (a)~$(\theta_a,\theta_b)=(20^\circ,-28^\circ)$ metasurface. (b)~$(\theta_a,\theta_b)=(0^\circ,-70^\circ)$ metasurface.}\label{fig:ms_picture}
\end{figure}

\begin{figure}[h]
\centering
\includegraphics[width=0.9
\columnwidth]{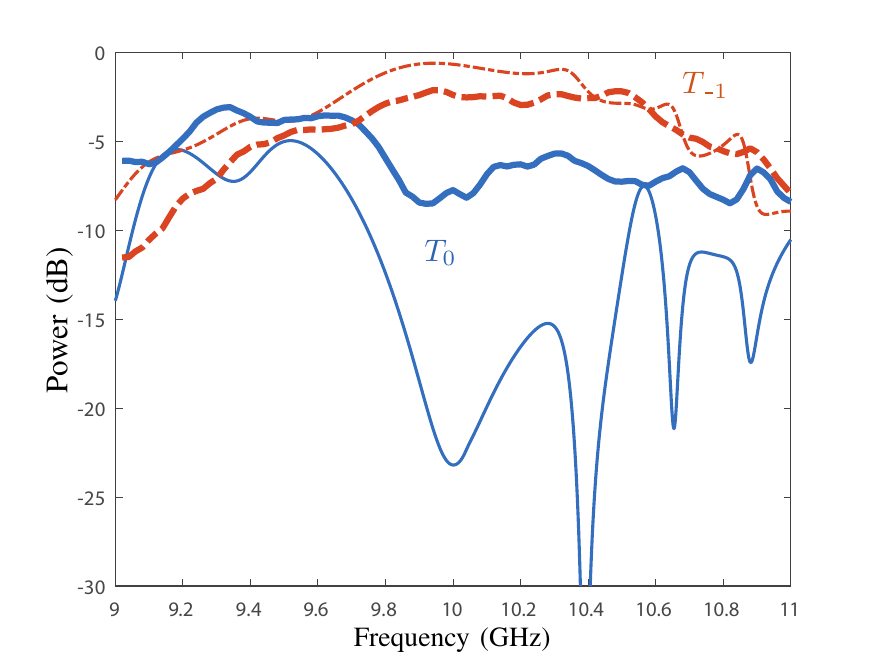}
\caption{Measured (thick lines) and simulated (thin lines) scattering parameters in transmission of the $(\theta_a,\theta_b)=(20^\circ,-28^\circ)$ metasurface.}\label{fig:ms60_exp}
\end{figure}

\begin{figure}[h]
\centering
\includegraphics[width=0.9
\columnwidth]{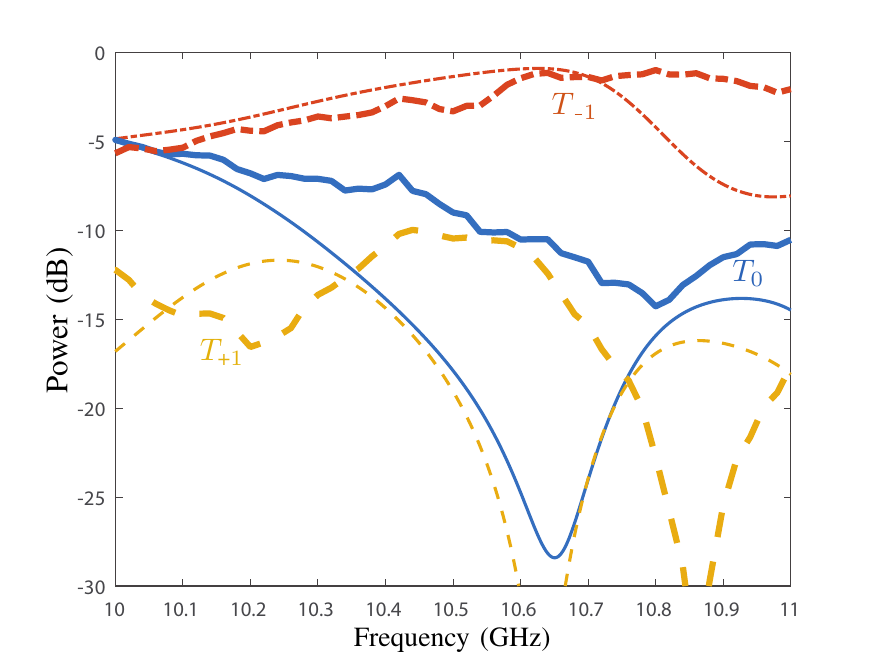}
\caption{Measured (thick lines) and simulated (thin lines) scattering parameters in transmission of the $(\theta_a,\theta_b)=(0^\circ,-70^\circ)$ metasurface. Note that the optimal frequency after fabrication was up-shifted by about 0.2~GHz compared to the simulation.}\label{fig:ms80_exp}
\end{figure}

The two metasurfaces were fabricated and measured. Figure~\ref{fig:ms_picture} shows a photograph of them.
In the measurement, we used a horn antenna placed $\sim400$~mm from the metasurface and a near-field probe scanning over a plane parallel to the metasurface in the transmission region. We then applied a near-field to far-field transformation~\cite{balanis2016antenna} to evaluate the transmission response of the metasurface. The measurement results are shown, superimposed with the simulations, in Figs.~\ref{fig:ms60_exp} and~\ref{fig:ms80_exp} for the $(\theta_a,\theta_b)=(20^\circ,-28^\circ)$ metasurface and  the $(\theta_a,\theta_b)=(0^\circ,-70^\circ)$ metasurface, respectively. The observed discrepancies between simulation and measurement may be attributed to different factors, including fabrication tolerance, horn antenna excitation~(instead of ideal plane wave) and probe antenna imperfection~(spurious edge diffraction).

The performance of our metasurfaces is limited only by dissipation loss in the metal scatterers and in the dielectric. In~\cite{selvanayagam2013discontinuous}, the authors established a theoretical limit in efficiency for a lossless monoanisotropic metasurface, which is found to be $\thicksim76\%$ for $(\theta_a,\theta_b)=(0^\circ,-70^\circ)$; it is interesting to note that our bianisotropic metasurface exceeds this lossless monoanisotropic limit by around $5\%$ despite the natural presence of loss. The experimental work in~\cite{chen2017experimental}, for a similar wide-angle refraction,  had lower efficiency due to higher scattering into other diffraction orders and higher absorbtion compared to our metasurface.


\section{Conclusion}\label{sec:concl}

We have derived the conditions for diffraction-free refraction in a metasurface using a medium-based approach based on Generalized Sheet Transition Conditions (GSTCs) and surface susceptibility tensors, and experimentally demonstrated two diffraction-free metasurfaces that are essentially lossless, passive, bianisotropic and reciprocal.

Following~\cite{asadchy2016perfect}, we have considered refractive metasurfaces possessing only \emph{transverse} susceptibility components. However, diffraction-less refraction might also be achieved by metasurfaces including \emph{normal} polarizabilities, which would lead to other possibilities than the three reported in~\cite{asadchy2016perfect}. However, solving the synthesis problem for metasurfaces with nonzero normal susceptibility components is not trivial since the corresponding GSTCs relations form a set of \emph{differential} equations instead of just algebraic equations. At this stage, the design of such structures remains an open avenue for further investigation.

\bibliographystyle{IEEEtran}
\bibliography{LIB}

\begin{thebibliography}{10}
\providecommand{\url}[1]{#1}
\csname url@samestyle\endcsname
\providecommand{\newblock}{\relax}
\providecommand{\bibinfo}[2]{#2}
\providecommand{\BIBentrySTDinterwordspacing}{\spaceskip=0pt\relax}
\providecommand{\BIBentryALTinterwordstretchfactor}{4}
\providecommand{\BIBentryALTinterwordspacing}{\spaceskip=\fontdimen2\font plus
\BIBentryALTinterwordstretchfactor\fontdimen3\font minus
  \fontdimen4\font\relax}
\providecommand{\BIBforeignlanguage}[2]{{%
\expandafter\ifx\csname l@#1\endcsname\relax
\typeout{** WARNING: IEEEtran.bst: No hyphenation pattern has been}%
\typeout{** loaded for the language `#1'. Using the pattern for}%
\typeout{** the default language instead.}%
\else
\language=\csname l@#1\endcsname
\fi
#2}}
\providecommand{\BIBdecl}{\relax}
\BIBdecl

\bibitem{glybovski2016metasurfaces}
S.~B. Glybovski, S.~A. Tretyakov, P.~A. Belov, Y.~S. Kivshar, and C.~R.
  Simovski, ``Metasurfaces: From microwaves to visible,'' \emph{Phys. Rep.},
  vol. 634, 2016.

\bibitem{urbas2016roadmap}
A.~M. Urbas, Z.~Jacob, L.~Dal~Negro, N.~Engheta, A.~Boardman, P.~Egan, A.~B.
  Khanikaev, V.~Menon, M.~Ferrera, N.~Kinsey \emph{et~al.}, ``Roadmap on
  optical metamaterials,'' \emph{J. Opt.}, vol.~18, no.~9, p. 093005, 2016.

\bibitem{tretyakov2015metasurfaces}
S.~Tretyakov, ``Metasurfaces for general transformations of electromagnetic
  fields,'' \emph{Phil. Trans. R. Soc. A}, vol. 373, no. 2049, p. 20140362,
  2015.

\bibitem{ra2013total}
Y.~Ra'di, V.~S. Asadchy, and S.~A. Tretyakov, ``Total absorption of
  electromagnetic waves in ultimately thin layers,'' \emph{IEEE Antenn.
  Wireless Propag. Lett.}, vol.~61, no.~9, pp. 4606--4614, 2013.

\bibitem{shi2014dual}
H.~Shi, A.~Zhang, S.~Zheng, J.~Li, and Y.~Jiang, ``Dual-band polarization angle
  independent 90 polarization rotator using twisted electric-field-coupled
  resonators,'' \emph{Appl. Phys. Lett.}, vol. 104, no.~3, p. 034102, 2014.

\bibitem{almoneef2015metamaterial}
T.~S. Almoneef and O.~M. Ramahi, ``Metamaterial electromagnetic energy
  harvester with near unity efficiency,'' \emph{Appl. Phys. Lett.}, vol. 106,
  no.~15, p. 153902, 2015.

\bibitem{pfeiffer2014controlling}
C.~Pfeiffer and A.~Grbic, ``Controlling vector {B}essel beams with
  metasurfaces,'' \emph{Phys. Rev. Applied}, vol.~2, no.~4, p. 044012, 2014.

\bibitem{achouri2016comparison}
K.~Achouri, G.~Lavigne, and C.~Caloz, ``Comparison of two synthesis methods for
  birefringent metasurfaces,'' \emph{J. Appl. Phys.}, vol. 120, no.~23, p.
  235305, 2016.

\bibitem{achouri2016metasurface}
K.~Achouri, G.~Lavigne, M.~A. Salem, and C.~Caloz, ``Metasurface spatial
  processor for electromagnetic remote control,'' \emph{IEEE Antenn. Wireless
  Propag. Lett.}, vol.~64, no.~5, pp. 1759--1767, 2016.

\bibitem{khorasaninejad2016metalenses}
M.~Khorasaninejad, W.~T. Chen, R.~C. Devlin, J.~Oh, A.~Y. Zhu, and F.~Capasso,
  ``Metalenses at visible wavelengths: Diffraction-limited focusing and
  subwavelength resolution imaging,'' \emph{{S}cience}, vol. 352, no. 6290, pp.
  1190--1194, 2016.

\bibitem{yu2011light}
N.~Yu, P.~Genevet, M.~A. Kats, F.~Aieta, J.-P. Tetienne, F.~Capasso, and
  Z.~Gaburro, ``Light propagation with phase discontinuities: generalized laws
  of reflection and refraction,'' \emph{{S}cience}, vol. 334, no. 6054, pp.
  333--337, 2011.

\bibitem{neviere1991electromagnetic}
M.~Neviere, ``Electromagnetic study of transmission gratings,'' \emph{Appl.
  Opt.}, vol.~30, no.~31, pp. 4540--4547, 1991.

\bibitem{destouches200599}
N.~Destouches, A.~Tishchenko, J.~Pommier, S.~Reynaud, O.~Parriaux, S.~Tonchev,
  and M.~A. Ahmed, ``99\% efficiency measured in the-1 st order of a resonant
  grating,'' \emph{Opt. Express}, vol.~13, no.~9, pp. 3230--3235, 2005.

\bibitem{asadchy2016perfect}
V.~S. Asadchy, M.~Albooyeh, S.~N. Tcvetkova, A.~D\'{\i}az-Rubio, Y.~Ra'di, and
  S.~A. Tretyakov, ``Perfect control of reflection and refraction using
  spatially dispersive metasurfaces,'' \emph{Phys. Rev. B}, vol.~94, p. 075142,
  Aug 2016.

\bibitem{estakhri2016wave}
N.~M. Estakhri and A.~Al{\`u}, ``Wave-front transformation with gradient
  metasurfaces,'' \emph{Phys. Rev. X}, vol.~6, no.~4, p. 041008, 2016.

\bibitem{epstein2016arbitrary}
A.~Epstein and G.~V. Eleftheriades, ``Arbitrary power-conserving field
  transformations with passive lossless omega-type bianisotropic
  metasurfaces,'' \emph{IEEE Trans. Antennas Propag.}, vol.~64, pp. 3880 --
  3895, 2016.

\bibitem{achouri2014general}
K.~Achouri, M.~A. Salem, and C.~Caloz, ``General metasurface synthesis based on
  susceptibility tensors,'' \emph{IEEE Trans. Antennas Propag.}, vol.~63,
  no.~7, pp. 2977--2991, July 2015.

\bibitem{zhu2015passive}
B.~O. Zhu and Y.~Feng, ``Passive metasurface for reflectionless and arbitary
  control of electromagnetic wave transmission,'' \emph{IEEE Trans. Antennas
  Propag.}, vol.~63, no.~12, pp. 5500--5511, 2015.

\bibitem{wong2016reflectionless}
J.~P. Wong, A.~Epstein, and G.~V. Eleftheriades, ``Reflectionless wide-angle
  refracting metasurfaces,'' \emph{IEEE Antenn. Wireless Propag. Lett.},
  vol.~15, pp. 1293--1296, 2016.

\bibitem{chen2017experimental}
M.~Chen, E.~Abdo-S{\'a}nchez, A.~Epstein, and G.~V. Eleftheriades,
  ``Experimental verification of reflectionless wide-angle refraction via a
  bianisotropic {H}uygens' metasurface,'' \emph{arXiv preprint
  arXiv:1703.06669}, 2017.

\bibitem{lavigne2017perfectly}
G.~Lavigne, K.~Achouri, C.~Caloz, V.~Asadchy, and S.~Tretyakov, ``Perfectly
  refractive metasurface using bianisotropy,'' \emph{arXiv preprint
  arXiv:1704.01641}, 2017.

\bibitem{lax1962microwave}
B.~Lax and K.~J. Button, ``Microwave ferrites and ferrimagnetics,'' 1962.

\bibitem{kodera2011artificial}
T.~Kodera, D.~L. Sounas, and C.~Caloz, ``Artificial {F}araday rotation using a
  ring metamaterial structure without static magnetic field,'' \emph{Appl.
  Phys. Lett.}, vol.~99, no.~3, p. 031114, 2011.

\bibitem{wang2012gyrotropic}
Z.~Wang, Z.~Wang, J.~Wang, B.~Zhang, J.~Huangfu, J.~D. Joannopoulos,
  M.~Solja{\v{c}}i{\'c}, and L.~Ran, ``Gyrotropic response in the absence of a
  bias field,'' \emph{Proc. Natl. Acad. Sci. U.S.A.}, vol. 109, no.~33, pp.
  13\,194--13\,197, 2012.

\bibitem{taravati2017nonreciprocal}
S.~Taravati, B.~A. Khan, S.~Gupta, K.~Achouri, and C.~Caloz, ``Nonreciprocal
  nongyrotropic magnetless metasurface,'' \emph{IEEE Trans. Antennas Propag.},
  2017.

\bibitem{achouri2015synthesis}
K.~Achouri, B.~A. Khan, S.~Gupta, G.~Lavigne, M.~A. Salem, and C.~Caloz,
  ``Synthesis of electromagnetic metasurfaces: principles and illustrations,''
  \emph{EPJ Appl. Metamater.}, vol.~2, p.~12, 2015.

\bibitem{achouri2017mathematical}
K.~Achouri, Y.~Vahabzadeh, and C.~Caloz, ``Mathematical synthesis and analysis
  of nonlinear metasurfaces,'' \emph{arXiv preprint arXiv:1703.09082}, 2017.

\bibitem{idemen2011discontinuities}
M.~M. Idemen, \emph{Discontinuities in the electromagnetic field}.\hskip 1em
  plus 0.5em minus 0.4em\relax John Wiley \& Sons, 2011, vol.~40.

\bibitem{rothwell2008electromagnetics}
E.~J. Rothwell and M.~J. Cloud, \emph{Electromagnetics}.\hskip 1em plus 0.5em
  minus 0.4em\relax CRC press, 2008.

\bibitem{salem2014manipulating}
M.~A. Salem and C.~Caloz, ``Manipulating light at distance by a metasurface
  using momentum transformation,'' \emph{Opt. Express}, vol.~22, no.~12, pp.
  14\,530--14\,543, 2014.

\bibitem{ishimaru1991electromagnetic}
A.~Ishimaru, \emph{Electromagnetic wave propagation, radiation, and
  scattering}.\hskip 1em plus 0.5em minus 0.4em\relax Prentice-Hall, 1991.

\bibitem{balanis2016antenna}
C.~A. Balanis, \emph{Antenna theory: analysis and design}.\hskip 1em plus 0.5em
  minus 0.4em\relax John Wiley \& Sons, 2016.

\bibitem{selvanayagam2013discontinuous}
M.~Selvanayagam and G.~V. Eleftheriades, ``Discontinuous electromagnetic fields
  using orthogonal electric and magnetic currents for wavefront manipulation,''
  \emph{Opt. Express}, vol.~21, no.~12, pp. 14\,409--14\,429, 2013.

\end{thebibliography}

\end{document}